\documentclass[%
 reprint,
showkey,
 amsmath,amssymb,
 aps,
prd,
]{revtex4-1}

\usepackage{graphicx}
\usepackage{dcolumn}
\usepackage{bm}
\usepackage{slashed}
\usepackage{color}
\usepackage{textcomp}

\def\baas{Bulletin of the American Astronomical Society}
\def\ssr{Space Sci. Rev.}
\def\mnras{Mon. Not. R. Astron. Soc.}%
\def\apj{Astrophys. J.}
\def\apjl{Astrophys. J. Lett.}
\def\prl{Phys. Rev. Lett.}
\def\prd{Phys. Rev. D}
\def\pr{Phys. Rev.}
\def\aap{Astron. Astrophys.}

\def\nat{Nature}

\begin{document}

\preprint{APS/123-QED}

\title{Retrieving the True Masses of Gravitational-wave Sources}

\author{Xian Chen}
\email{Corresponding author: xian.chen@pku.edu.cn}
\affiliation{Astronomy Department, School of Physics, Peking University, 100871 Beijing, China}
\affiliation{Kavli Institute for Astronomy and Astrophysics at Peking University, 100871 Beijing, China}

\author{Zhe-Feng Shen}
\affiliation{Astronomy Department, School of Physics, Peking University, 100871 Beijing, China}

\date{\today}

\begin{abstract}
Gravitational waves (GWs) encode important information about the mass of the
source. For binary black holes (BBHs), the templates that are used to retrieve
the masses normally are developed under the assumption of a vacuum environment.
However, theories suggest that some BBHs form in gas-rich environments. Here we
study the effect of hydrodynamic drag on the chirp signal of a stellar-mass BBH
and the impact on the measurement of the mass.  Based on theoretical arguments,
we show that the waveform of a BBH in gas resembles that of a more massive BBH
residing in a vacuum. The effect is important for LISA sources but negligible
for LIGO/Virgo binaries.  Furthermore, we carry out a matched-filtering search
of the best fitting parameters. We find that the best-fit chirp mass could be
significantly greater than the real mass if the gas effect is not appropriately
accounted for.  Our results have important implications for the future joint
observation of BBHs using both ground- and space-based detectors.
\end{abstract}

\maketitle

\section{Introduction}

Measuring the mass of a gravitational wave (GW) source is an old but difficult
problem because the observable is not mass, but the phase and amplitude of GWs.
A model is needed to translate the observables to the mass. However, the
standard model that is used in the current GW observations often neglects the
astrophysical factors which could affect the dynamical evolution of the source
or the propagation of the GWs.  Consequently, the standard model could
potentially misinterpret the GW signal. 

For example, redshift is such a factor \cite{schutz86,chen19}.  It stretches
the waveform so that a low-mass source at high redshift looks identical to a
massive one at low redshift. Such a ``mass-redshift'' degeneracy has become a
serous issue because the Laser Interferometer Gravitational-wave Observatory
(LIGO) and the Virgo detectors have detected seemly over-massive black holes
(BHs) \cite{ligo_2018a,ligo_2018b}, which are two to three times bigger than
those BHs previously detected in X-ray binaries \cite{mcclintock14,corral16}.
On one hand, the high masses may be real and reflect the peculiarity of the
environment in which the BHs form
\cite{ligo16astro,ama16,stone17,bartos17,mckernan18,secunda19}.  Alternative,
the BHs may be intrinsically small but appear more massive due to a high
redshift.

The high redshift could be explained in two astrophysical scenarios.  One
possibility is that binary BHs (BBHs) coalesce at high cosmological redshift and get
strongly lensed by foreground galaxies or galaxy clusters
\cite{Broadhurst18,smith18}. This scenario, although possible for a small
number of of BBHs, could not account for all the massive BHs detected so far by
LIGO/Virgo \cite{ligo_2018a,hannuksela19}. Another possibility is that BBHs
merge in places very close to supermassive black holes (SMBHs), so that both
the Doppler and gravitational redshift become significant
\cite{chen18,han19,chen19}. The problem with this scenario is that the event
rate is difficult to estimate because the stellar distribution
around SMBHs is not well constrained by observations \cite{chen19}. 

Is there a way of distinguishing, on a one-to-one basis, the redshifted BBHs
from the intrinsically massive ones? It is difficult using ground-based
detectors. The corresponding signals usually last no more than one second, too
short to reveal any signature of gravitational lensing or a nearby SMBH.
However, with a space-borne GW detector, such as the Laser Interferometer Space
Antenna (LISA \cite{danzmann_2017}), the answer would be different.  Being
sensitive to milli-Hertz (mHz) GWs, LISA could detect BBHs at a much earlier
evolutionary stage, weeks to millenniums before they enter the LIGO/Virgo band
\cite{miller02,sesana16,chen17}. The corresponding signals could be as long as
the lifetime of LISA, about $4-5$ years. Earlier studies showed that if a BBH
is strongly lensed, LISA could detect multiple images of the source
\cite{seto04,sereno11} or, in some rare cases, detect a shift of the GW phase
caused by the wave effect of gravitational lensing
\cite{nakamura98,takahashi03lensing}. If, on the other hand, a BBH is close to
a SMBH, LISA could detect a distortion of the waveform caused by either the
orbital motion of the binary around the SMBH
\cite{inayoshi17,meiron17,robson18,chamberlain_moore_2019,wong19} or the tidal
force of the SMBH \cite{meiron17,hoang19,randall19,fang19}.

Besides redshift, are there other astrophysical factors which could affect the
measurement of the masses of GW sources? In this article, we show that the
presence of gas around BBHs could also lead to an overestimation of the masses.
Investigating this scenario is important because in many theoretical models the
merger of stellar-mass BBHs is driven by gas (e.g.
\cite{stone17,bartos17,secunda18,mckernan18,antoni19,loeb16,macleod17,dorazio18,tagawa18}).

\section{The effect of gas on the measurement of mass}\label{sec:theory}

The waveform of a merging BBH can be divided into three parts, corresponding to the
inspiral, merger, and ringdown phases
\cite{centrella10}.
During the inspiral phase, the information of mass is encoded in the GW frequency ($f$)
and the time derivative of it ($\dot{f}$). For example, consider a BBH whose BH
masses are $m_1$ and $m_2$ (we assume $m_1\ge m_2$).  If the binary is in
vacuum, the following quantity
\begin{equation}
{\cal M}:=\frac{c^3}{G}\left(\frac{5f^{-11/3}\dot{f}}{96\pi^{8/3}}\right)^{3/5}\label{eq:Mc}
\end{equation}
is equivalent to
\begin{equation}
{\cal M}=\frac{(m_1m_2)^{3/5}}{(m_1+m_2)^{1/5}}
\end{equation}
according to the Newtonian approximation \cite{cutler94}, where $G$ is the
gravitational constant and $c$ the speed of light.  This quantity has the
dimension of mass and uniquely determines how the GW frequency increases with
time. It is known as the ``chirp mass''. 

Gas could make BBHs shrink more rapidly by imposing a hydrodynamical drag on
each BH \cite{ostriker99}. As a result, the observed $\dot{f}$ would be bigger
than that in the vacuum case. Without knowing the gas effect, an observer is
likely to apply the vacuum model, i.e.,  Equation~(\ref{eq:Mc}), to the
observed $f$ and $\dot{f}$.  The derived mass, which we call ${\cal M}_o$,
will be bigger than the real chirp mass. 

To see this effect more clearly, we first denote the value of $\dot{f}$ in the
vacuum model as $\dot{f}_{\rm gw}$ and that in the gas model as $\dot{f}_{\rm
gas}$.  Furthermore, we define the semi-major axis of a BBH as $a$ and the shrinking rate
due to GW radiation as $\dot{a}_{\rm gw}$. Then we can express the shrinking
timescale due to GW radiation as $T_{\rm gw}:=a/|\dot{a}_{\rm gw}|$.  Now
suppose the presence of gas causes the binary to shrink at an additional rate
of $\dot{a}_{\rm gas}$, we can write the gas-drag timescale as $T_{\rm
gas}:=a/|\dot{a}_{\rm gas}|$. Let us further assume for simplicity a circular
binary, so that the GW frequency is $f=\pi^{-1}\sqrt{G(m_1+m_2)/a^3}$.  From
the last formula we find $\dot{f}_{\rm gas}=(1+T_{\rm gw}/T_{\rm
gas})\dot{f}_{\rm gw}$.  Substituting $\dot{f}_{\rm gas}$ in
Equation~(\ref{eq:Mc}) for $\dot{f}$, we find that the ``observable'' mass  is
no long the intrinsic chirp mass, but
\begin{equation}
{\cal M}_o=(1+T_{\rm gw}/T_{\rm gas})^{3/5}{\cal M}.\label{eq:Mbig}
\end{equation}
Interestingly, this mass could be much bigger than the real mass when $T_{\rm gas}\ll T_{\rm gw}$.

Now we compare the values of $T_{\rm gas}$ and $T_{\rm gw}$.  
For circular binary and in the Newtonian approximation, the GW radiation
timescale can be calculated with
\begin{align}
T_{\rm gw}&:=\frac{a}{|\dot{a}_{\rm gw}|}=\frac{5}{64}\frac{c^5a^4}{G^3m_1m_2m_{12}}\\
&\simeq\frac{9.1\times10^3}{q(1+q)^{-1/3}}
\left(\frac{m_1}{10\,M_\odot}\right)^{-5/3}
\left(\frac{f}{3\,{\rm mHz}}\right)^{-8/3}\,{\rm years}\label{eq:Tgw}
\end{align}
(from \cite{peters64}), where $q:=m_1/m_1$ is the mass ratio of the binary and $m_{12}:=m_1+m_2$.
We are scaling the GW frequency to mHz because the corresponding semi-major axis is about 
\begin{align}
a&=\left(\frac{Gm_{12}}{\pi^2 f^2}\right)^{1/3}\\
&\simeq0.0021
\left(\frac{m_{12}}{20~M_\odot}\right)^{1/3}
\left(\frac{f}{3\,{\rm mHz}}\right)^{-2/3}\,{\rm AU}.
\end{align}
BBHs with such a semi-major axis could have a gas-drag timescale as short as
$T_{\rm gas}\simeq10^3$ years according to the earlier studies of the BBHs in
gaseous environments (e.g. \cite{bartos17}).  It is worth noting that $T_{\rm
gas}$ is a function of gas density and hence could be even shorter in the most
gas-rich environment, such as the innermost part of the accretion disk around a
SMBH or the common envelope surrounding a binary \cite{antoni19,secunda19}.
From the timescales derived above, we find that for LISA BBHs ($f\sim1$ mHz) it
is possible that $T_{\rm gas}\ll T_{\rm gw}$.  For LIGO/Virgo BBHs ($f\sim10$
Hz), gas drag is no longer important because the GW radiation timescale,
according to Equation~(\ref{eq:Tgw}), is too short.

To be more quantitative, take $T_{\rm gas}=10^3$ years and $T_{\rm gw}=10^4$
years for example.  We have $T_{\rm gw}/T_{\rm gas}=10$. According to
Equation~(\ref{eq:Mbig}) one would overestimate the mass by a factor of $4.2$
if the gas effect is ignored.  In this case, a BBH with $m_1=m_2=10\,M_\odot$
(${\cal M}\simeq8.7\,M_\odot$) would appear to have a chirp mass of ${\cal
M}_o\simeq37\,M_\odot$. In other words, from LISA waveform it seems that two
$42\,M_\odot$ BHs are merging.

\section{Matched filtering and parameter estimation}

In practice, LISA employs a technique called the ``matched filtering'' to estimate the
parameters of a GW source \cite{finn92}. In this technique, the similarity of two waveforms, say
$h_1(t)$ and $h_2(t)$, is quantified by the ``fitting factor'' (FF), defined as
\begin{equation}
{\rm FF}=\frac{\left<h_1|h_2\right>}{\sqrt{\left<h_1|h_1\right>\left<h_2|h_2\right>}},
\end{equation}
where $\left<h_1|h_2\right>$ denotes an inner product, which can be calculated with
\begin{equation}
\left<h_1|h_2\right>=2\int_0^\infty\frac{\tilde{h}_1(f)\tilde{h}_2^*(f)+\tilde{h}_1^*(f)\tilde{h}_2}{S_n(f)}df.
\end{equation}
In the last equation, the tilde symbols stand for the Fourier transformation
and the stars stand for the complex conjugation. The quantity $S_n(f)$ is the
spectral noise density of LISA (see details in \cite{han19}). An exact match,
in principle, would mean ${\rm FF}=1$. 

In reality, noise exists and consequently  FF is not unity even when $h_1$ and
$h_2$ are identical. A more practical definition of ``match'' is
that $\left<\delta h|\delta h\right> <1$, where $\delta
h:=\tilde{h}_1(f)-\tilde{h}_2(f)$ \cite{lindblom08}.  There is a close
relationship between $\left<\delta h|\delta h\right>$, FF, and the
signal-to-noise ratio (SNR) defined as $\rho^2:=\left<h|h\right>$, which could help
simplifying our calculation. We note that in GW observations often we are in
the situation where $h_1\simeq h_2$. Therefore, we have
$\rho^2\simeq\left<h_1|h_1\right>\simeq\left<h_2|h_2\right>$. Using the last
equation, we find that the condition for match is equivalent to 
\begin{equation}
{\rm
FF}>1-1/(2\rho^2).\label{eq:FF}
\end{equation} 
Since LISA will claim a detection when $\rho\simeq10$
\cite{sesana16,kyutoku16}, only those temples with $FF>0.995$ are acceptable.

In our particular problem, $h_1(t)$ is the GW signal from a BBH embedded in a
gaseous environment, and $h_2(t)$ is the waveform template that we use to match
with $h_1(t)$ and extract physical parameters.  To prepare a template bank for
$h_2(t)$, a model is needed. So far, only the vacuum model has been considered
in the literature. In the following, we show that even though the vacuum model
is a wrong one, the resulting FF could still be very high. Consequently, using
this model will confuse the estimation of the mass of a BBH. 
 
In this work, we compute the waveforms using 
\begin{equation} h(t)=\frac{Am_1m_2}{a(t)}\cos\,\phi(t), \end{equation}
where $A$ is a normalization factor depending on the source distance but not important
for matched filtering, and
\begin{equation} \phi(t)=\int_0^{t}2\pi f(t')dt'+\phi_c \end{equation}
is the phase of GW. For $h_2(t)$, i.e., the vacuum model, $a(t)$ and $f(t)$ are
computed  following a post-Newtonian approximation (\cite{cutler94}).  For
$h_1(t)$, i.e., when there is gas, we add a term $1.5f/T_{\rm gas}$ to the
equation of $\dot{f}$ to mimic the effect of gas drag. 

To find the maximum FF, we explore the parameter space of $m_1$ and $\phi_c$
while keeping $q$ fixed to $0.7$, for simplicity. In a future work, we will
complete the analysis by searching in the full parameter space of
$(m_1,\,q,\,\phi_c)$.  Our fiducial parameters are ${\cal M}=8.7\,M_\odot$,
$q=0.7$, $a(0)=0.002$ AU, and $T_{\rm gas}\simeq10^3$ years.  The values of the first
three parameters are chosen such that in vacuum the BBH would merge on a
timescale of $T_{\rm gw}\simeq10^4$ years.

\begin{figure}[]
\centering
\includegraphics[width=0.5\textwidth]{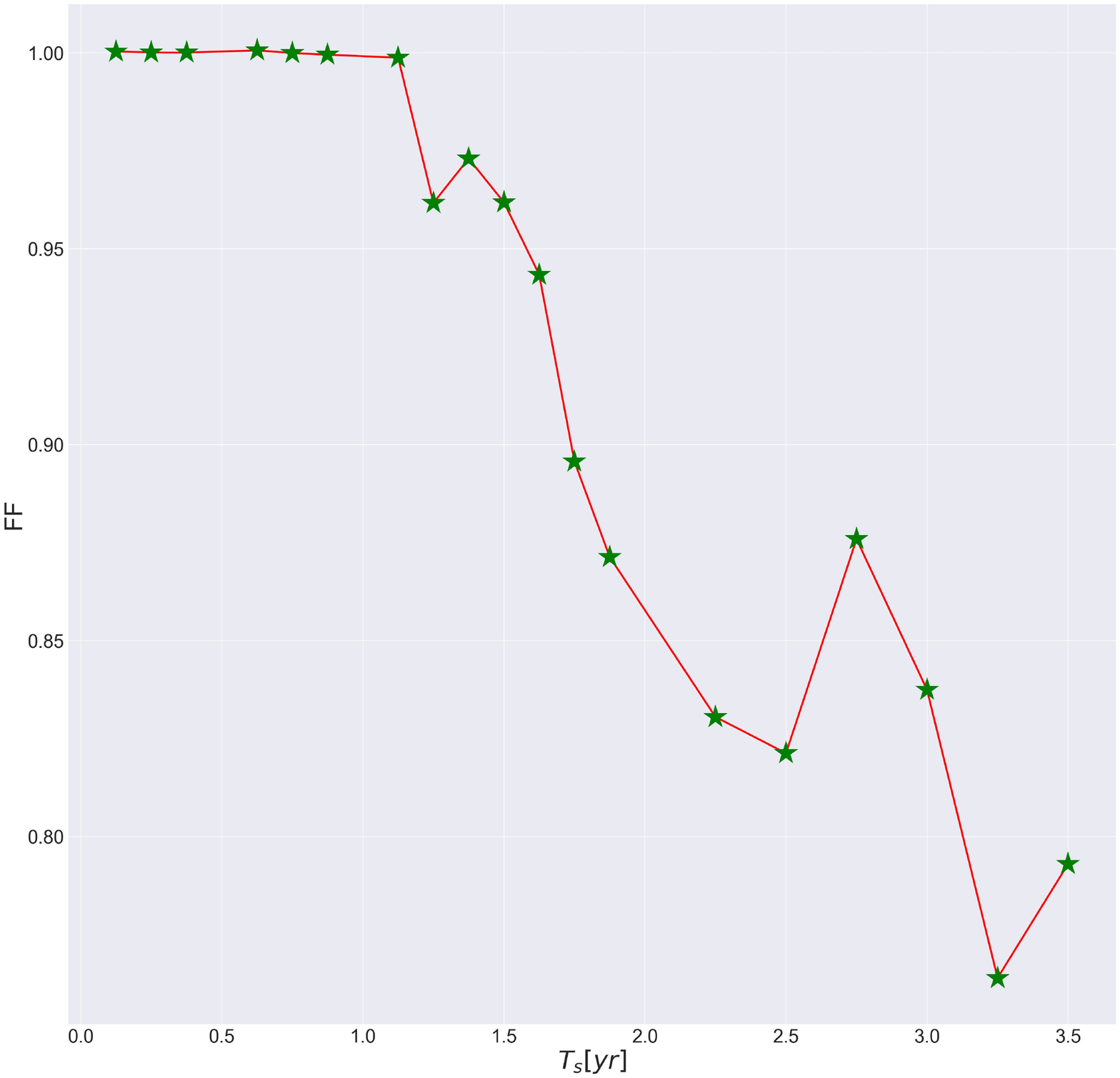}
\caption{Fitting factor as a function of the observing time of GWs.  The two
waveforms which are compared here are $h_1$, the waveform of a low-mass BBH
residing in a gaseous environment, and $h_2$, a high-mass BBH in vacuum.  The
input parameters are ${\cal M}=8.7\,M_\odot$, $q=0.7$, $a=0.002$ AU, and
$T_{\rm gas}\simeq10^3$ years.  The best-fit chirp mass, corresponding to the
highest FF, is ${\cal M}_o\simeq36.94\,M_\odot$.}\label{fig:FF} \end{figure}

Figure~\ref{fig:FF} shows the resulting FF as a function of the LISA observing
time, $T_s$. We find that FF is above $0.995$ during the first $1.1$ years of
observation.  The high FF means that the vacuum template gives a reasonable fit to
the signal, even though it is the wrong template to use here.  The best-fit
${\cal M}_o$ is about $36.94\,M_\odot$, much larger than ${\cal M}$.  This result
agrees well with what we have envisioned in Section~\ref{sec:theory}. 

When the observing time is longer than $1.1$ years, the FF decays to a value
below $0.995$.  This result indicates that LISA would be able to distinguish
the BBHs in gaseous environments from those in vacuum, given that the observing
time is long enough. The exact time that is needed to reveal the
difference depends on the parameters of the BBH, as well as the properties of
the surrounding gas. This issue deserves further investigation.

\section{Conclusions}

We have shown that the presence of gas around BBHs could affect the chirp
signal and lead to a significant overestimation of the mass of the binaries.
This effect is important for LISA observation but negligible for LIGO/Virgo
sources.  Our results have important implications for the future joint
observation of BBHs using both ground and space-borne detectors
\cite{sesana17,cutler19}.

\begin{acknowledgments}
This research was funded by the ``985 Project'' of Peking University
and the National Science Foundation of China No. 11873022. XC is partly
supported by the Strategic Priority Research Program of the Chinese Academy of
Sciences, Grant No. XDB23040100 and No. XDB23010200. The matched filtering is
performed on the High-Performance Computing
Platform of Peking University. 
\end{acknowledgments}

\bibliographystyle{apsrev4-1.bst}

\end{document}